\documentclass[11pt,a4paper]{article}
\pdfoutput=1
\usepackage{jheppub}
\usepackage[T1]{fontenc}
\usepackage{graphicx} 
\usepackage{dcolumn}
\usepackage{bm}
\usepackage{amssymb}
\usepackage{amsmath,amsthm}
\usepackage{bm}
\usepackage{mathrsfs}

\usepackage{slashed}

\title{$\mathcal{O}(\alpha^2)$ ISR effects with a full electroweak one-loop correction for a top pair-production at the ILC}
\author[a]{Junpei Fujimoto}
\author[a,1]{Yoshimasa Kurihara\note{Corresponding author.}}
\author[a,b]{Nhi M. U. Quach}

\affiliation[a]{High Energy Accelerator Research Organization (KEK), Tsukuba, Ibaraki 305-0801, Japan. }
\affiliation[b]{The Graduate University for Advanced Studies (SOKENDAI), Hayama, Kanagawa 240-0193, Japan.}

\emailAdd{junpei@post.kek.jp}
\emailAdd{yoshimasa.kurihara@kek.jp}
\emailAdd{nhiquach@post.kek.jp}
\abstract{
Precise predictions for an $e^+e^-\rightarrow t\bar{t}$ cross section are presented at an energy region from 400 GeV to 800 GeV.
Cross sections are estimated including the beam-polarization effects with full $\mathcal{O}(\alpha)$, and also with effects of  the initial-state photon emission.
A radiator technique is used for the initial-state photon emission up to two-loop order.
A weak correction is defined as the full electroweak corrections without the initial-state photonic corrections.
As a result, it is obtained that the total cross section of  a top quark pair-production receives  the weak corrections of $+4\%$ over the trivial initial state corrections at a centre of mass energy of 500 GeV. 
Among the initial state contributions, a  contribution from two-loop diagrams gives less than $0.11\%$ correction over the one-loop ones at the center of mass energies of from $400$ GeV to $800$ GeV.
In addition, an effect of a running coupling constant is also discussed.
}
\begin{document} 
\maketitle
\flushbottom
%
%
\section{Introduction}
The standard theory of particle physics are established finally by a discovery of the Higgs boson\cite{Aad:2012tfa,Chatrchyan201230} in 2012.
A current target of particle physics is searching for a more fundamental theory beyond the standard model (BSM).
A keystone along this direction must be the Higgs boson and top quark.
Since the top quark is the heaviest fermion with a mass at the electroweak symmetry-breaking scale, it is naturally expected to have a special role in the BSM\@. 

The international linear collider\cite{Behnke:2013xla} (ILC), which is an electron-positron colliding experiment with centre of mass (CM) energies above 250 GeV, is proposed and intensively discussed  as a future project of high-energy physics. 
One of main goals of ILC experiments is a precise measurement of top quark properties.
Detailed Monte Carlo studies have shown that the ILC would be able to measure most of the standard model parameters to within sub-percent levels\cite{Baer:2013cma}.
Because of the improvement of the experimental accuracy of the ILC, theoretical predictions are required be given with new level of precision.
In particular, a radiative correction due to the electroweak interaction (including spin polarizations) is mandatory for such requirements.
Before the discovery of the top quark, a full electroweak radiative correction was conducted for an $e^-e^+\rightarrow t\bar{t}$ process at a lower energy\cite{Fujimoto:1987hu}, and was then obtained independently for higher energies\cite{Fleischer:2002rn,Fleischer2003}.
The same correction fo a $e^-e^+\rightarrow t\bar{t}\gamma$ process has also been reported\cite{Khiem2013}.
Recently, full electroweak radiative corrections for the process 
$e^-e^+\rightarrow t \bar{t}\rightarrow b \bar{b} \mu^+\mu^-\nu_{\mu} \bar{\nu}_{\mu}$
using a narrow-width approximation for the  top quarks including the spin-polarization effects are reported\cite{Quach:2017ijt} by authors including those of a present report.
A possible search of the minimum SUSY particles trough loop corrections of the top-quark pair-production at the ILC is reported in \cite{doi:10.1093/ptep/ptx048}.

Among several sources of the radiative correction, it is known that the initial state photonic correction gives the largest contribution in general, and thus it is important for a precise estimation of production cross sections.
In this report, the precise estimation of the effect due to the initial-state photon-radiations is discussed in detail 

%
%
\section{Calculation Method}\label{CM}
\subsection{GRACE system}
For precise cross-section calculations of the target process,  a GRACE-Loop system is used in this study.
The GRACE system is an automatic system to calculate cross sections of scattering processes at one-loop level for the standard theory\cite{Belanger2006117} and the minimum SUSY model\cite{PhysRevD.75.113002}. 
The GRACE system has treated  electroweak processes with two, three or four particles in the final state are calculated\cite{Belanger2003152,Belanger2003163, Belanger2003353,Kato:2005iw} at the one-loop order.
The renormalization of the electroweak interaction is carried out using on-shell scheme\cite{doi:10.1143/PTPS.73.1,doi:10.1143/PTPS.100.1}. 
Infrared divergences are regulated using fictitious photon-mass\cite{doi:10.1143/PTPS.100.1}.
The symbolic manipulation package FORM\cite{Vermaseren:2000nd} is used to handle all Dirac and tensor algebra in $n$-dimensions. 
For loop integrations, all tensor one-loop integrals are reduced to scalar integrals using our own formalism\cite{Belanger2006117}, then performed integrations using packages FF\cite{vanOldenborgh:1990yc} or LoopTools\cite{Hahn:1998yk}. 
Phase-space integrations are done using an adaptive Monte Carlo integration package BASES\cite{Kawabata1986127,KAWABATA1995309}.
For numerical calculations, we use a quartic precision for floating variables. 

While using $R_\xi$-gauge in linear gauge-fixing terms in the GRACE system,, the non-linear gauge fixing Lagrangian\cite{Boudjema:1995cb,Belanger2006117} is also employing for the sake of the system checking.
Before calculating cross sections, we performed numerical tests to confirm that the amplitudes are independent of all redundant parameters around $20$ digits at several randomly chosen phase points.
In addition to above checks, soft-photon cut-off independence was examined:
cross sections at the one-loop level, results must be independent from a head-photon cut-off parameter $k_c$. 
We confirmed that, while varying a parameter $k_c$ from $10^{-4}$ GeV to $10^{-1}$ GeV,  the results of numerical phase-space integrations are consistent  each other within the statistical errors of numerical integrations, that is typically around $0.1\%$ order.

\subsection{Radiator method}
The effect of the initial photon emission can be factorized when a total energy of emitted photons are small enough compared with a beam energy or a small angle (co-linear) emission .
The calculations under a such approximation is referred as to the ``soft-colinear photon approximation(SPA)''.
Under the SPA, the corrected cross sections with the initial state photon radiation(ISR), $\sigma_{ISR}$, can be obtained from the tree cross sections $\sigma_{Tree}$ using a structure function $H(x,s)$ as follows:
\begin{eqnarray} 
\sigma_{ISR}&=&\int^1_0 dx~H(x,s)\sigma_{Tree}\left(s(1-x)\right),\label{ISRTree}
\end{eqnarray}
where $s$ is the CM energy square and $x$ is a energy fraction of an emitted photon.
The structure function can be calculated using the perturbative method with the SPA.
Concrete formulae of the structure function are calculated up to two loop order\cite{doi:10.1143/PTPS.100.1}. 
A further improvement of the cross section estimation is possible using the ``exponentiation method''.
Initial state photon emissions under the SPA, a probability to emit each photon should be independent each other.
Thus, the probability to emitt any number of photons can be calculated as;
\begin{eqnarray} 
p&=&\sum_{k=0}^{\infty}p_k=\sum_{k=0}^{\infty}\frac{1}{k!}(p_1)^k,
\end{eqnarray}
where $p_k$ is a probability to emit $k$ photons. 
A factor $1/k!$ is necessary due to $k$ identical particles (photons) appearing in the final state.
This is nothing more than the Taylor expansion of the exponential function.
Therefore, the effect of the multiple photon emissions can be estimated by putting the one photon emission probability in an argument of the exponential function.
This technique is referred as to the exponentiation method.
 When the exponentiation method is applied to the cross section calculations at loop level, the corrected cross sections can not be expressed simply like the formula (\ref{ISRTree}), because the same loop corrections are included in both of the structure function and loop amplitudes. 
To avoid a double counting of the same corrections both in the structure function and loop amplitudes, one have to rearrange terms of corrections.

The total cross section at one-loop (fixed) order without the exponentiation, which is denoted as $\sigma_{NLO;fixed}$, can be expressed as;
\begin{eqnarray} 
\sigma_{NLO;fixed}&=&\sigma_{Loop}+\sigma_{Soft}+\sigma_{Hard}+\sigma_{Tree},\label{sNLOfinx}
\end{eqnarray}
where $\sigma_{Loop}$, $\sigma_{Soft}$ and $\sigma_{Hard}$ are the cross sections from the loop diagram, and soft and hard real-emission corrections, respectively.
A photon whose energy is greater (less) than the threshold energy $k_c$ is defined as a hard (soft) photon, respectively.
The soft-photon cross section can be expressed as $\sigma_{Soft}=\sigma_{Tree}\delta_{SPA}$, where a factorized function $\delta_{SPA}$ is obtained from the real-radiation diagrams using the SPA.
The SPA consists of three parts, the initial-state, final-state radiations and their interference terms. 
For the initial state radiation, it can be written as;
\begin{eqnarray}
\delta_{SPA}&=&\frac{\alpha}{\pi}\left(
2\left(L-1\right)\log{\frac{m_e}{\lambda}}+\frac{1}{2}L^2-2l\left(L-1\right)
\right),
\end{eqnarray}
where $\alpha$, $m_e$ and $\lambda$ are the fine structure constant, electron mass and fictitious photon mass, respectively.
Here two large log-factors appear as $L=\log{(s/m_e^2)}$ and $l=\log{(E/k_c)}$, where $E$ is a beam energy and $s=4E^2$.
The threshold energy $k_c$ is included in both $\sigma_{Hard}$ and $\delta_{SPA}$, and then the total cross section must be independent of the value of $k_c$ after summing up all contributions.
At the same time, final results are also independent from the photon mass $\lambda$ due to the cancellation among $\lambda$ contributions in $\sigma_{Loop}$ and $\delta_{SPA}$. 

The cross section with fixed order correction  (\ref{sNLOfinx}) can be improved using the exponentiation method.
To avoid a double counting of the terms appearing in both loop corrections and the structure function, terms must be re-arranged as follows:
\begin{eqnarray} 
\sigma_{NLO;ISR}&=&
\left(\sigma_{Loop}-\sigma_{Tree}\delta_{ISL}\right)+\widetilde{\sigma}_{Soft}
+\sigma_{Hard}+\widetilde{\sigma}_{ISR},\label{sNLOISR}
\end{eqnarray}
where $\delta_{ISL}$ is a correction factor from the initial state photon-loop diagrams, that can be given as;
\begin{eqnarray} 
\delta_{ISL}&=&\frac{2\alpha}{\pi}\left(
-\left(L-1\right)\log{\frac{m_e}{\lambda}}-\frac{1}{4}L^2+\frac{3}{4}L+\frac{\pi^2}{3}-1
\right).
\end{eqnarray}
This term must be subtracted from $\sigma_{Loop}$ because the same contribution is also included in the structure function $H(s,x)$.
A term $\widetilde{\sigma}_{Soft}$ includes only the final-state radiation and interference terms between the initial and final state radiations.
Instead, $\widetilde{\sigma}_{ISR}$ gives the improved cross section including the initial-state radiation using the radiator method\cite{doi:10.1143/PTPS.100.1}.
The total cross section can be calculated using the radiator function as;
\begin{eqnarray} 
\sigma_{ISR}&=&\int^{k^2_c/s}_0 dx_1 \int^{1-x1}_0dx_2~D(x_1,s)D(x_2,s)\sigma_{Tree}\left(sx_1x_2)\right).
\end{eqnarray}
The radiator function $D(x,s)$, which is corresponding to square root of the structure function,  gives a probability to emit a photon with energy fraction of $x$ at the CM energy square $s$.
In this method, electron and positron can emit different energies, and thus finite boost of the CM system can be treated.
The radiator function can be obtained as\cite{doi:10.1143/PTPS.100.1}.
\begin{eqnarray} 
D(1-x,s)^2&=&H(x,s)=\Delta_2\beta x^{\beta-1}
-\Delta_1\beta\left(1-\frac{x}{2}\right)\nonumber\\
&~&+\frac{\beta^2}{8}\left[
-4(2-x)\log{x}-\frac{1+3(1-x)^2}{x}\log{(1-x)}-2x
\right],\label{ISRLoop}
\end{eqnarray}
where
\begin{eqnarray*}
\beta&=&\frac{2\alpha}{\pi}\left(\log{\frac{s}{m_e^2}}-1\right),\\
\Delta_2=1+\delta_1+\delta_2,&~&\Delta_1=1+\delta_1\\
\delta_1=\frac{\alpha}{\pi}\left(\frac{3}{2}L+\frac{\pi^2}{3}-2\right),&~&
\delta_2=\left(\frac{\alpha L}{\pi}\right)^2
\left(
-\frac{1}{18}L+\frac{119}{72}-\frac{\pi^2}{3}
\right).
\end{eqnarray*}
In this case, the photon mass $\lambda$ is cancelled between $\sigma_{Loop}$ and $\sigma_{Tree}\delta_{ISL}$, and the threshold energy $k_c$ is cancelled between $\sigma_{Hard}$ and $\widetilde{\sigma}_{ISR}$.
This result is obtained based on perturbative calculations for initial-state photon emission diagrams up to two-loop order \cite{doi:10.1143/PTPS.100.1}.
Terms with $\alpha^2$ in (\ref{ISRLoop}) are obtained from two-loop diagrams.

%
%
\section{Results and discussions}\label{R&D}
\subsection{Input parameters}
The input parameters used in this report are listed in Table \ref{parameters}. 
The masses of the light quarks (i.e., other than the top quark) and $W$ boson are chosen to be consistent with low-energy experiments\cite{Khiem2015192}.
Other particle masses are taken from recent measurements\cite{Olive:2016xmw}.
The weak mixing-angle is given using the on-shell condition $\sin^2{\theta_W} = 1- m_W^2/m_Z^2$ because of our renormalization scheme.
The fine-structure constant $\alpha=1/137.0359859$ is taken from the low-energy limit of Thomson scattering, again because of our renormalization scheme.

Because the initial state photonic corrections are independent of the beam polarization, all cross section calculations in this report are performed with $100\%$ left (right) polarization of  an electron (positron), respectively.
	\begin{table}[b]
		\begin{center}
			\begin{tabular}{|c|c||c|c|}
\hline
$u$-quark mass & $58.0\times10^{-3}$ GeV & $d$-quark mass & $58.0\times10^{-3}$ GeV \\
$c$-quark mass & $1.5$ GeV & $s$-quark mass & $92.0\times10^{-3}$ GeV \\
$t$-quark mass & $173.5$ GeV & $b$-quark mass & $4.7$ GeV \\
$Z$-boson mass & $91.187$ GeV & $W$-boson mass & $80.370$ GeV \\
Higgs mass & 126 GeV & ~ & ~\\
\hline
			\end{tabular}
			\caption{Particle masses used in this report.}
\label{parameters}
		\end{center}
	\end{table}

\subsection{Electroweak radiative corrections}

\subsubsection{total cross sections}
\begin{figure}[t]
  	\begin{center}
  		\includegraphics[width={10cm}]{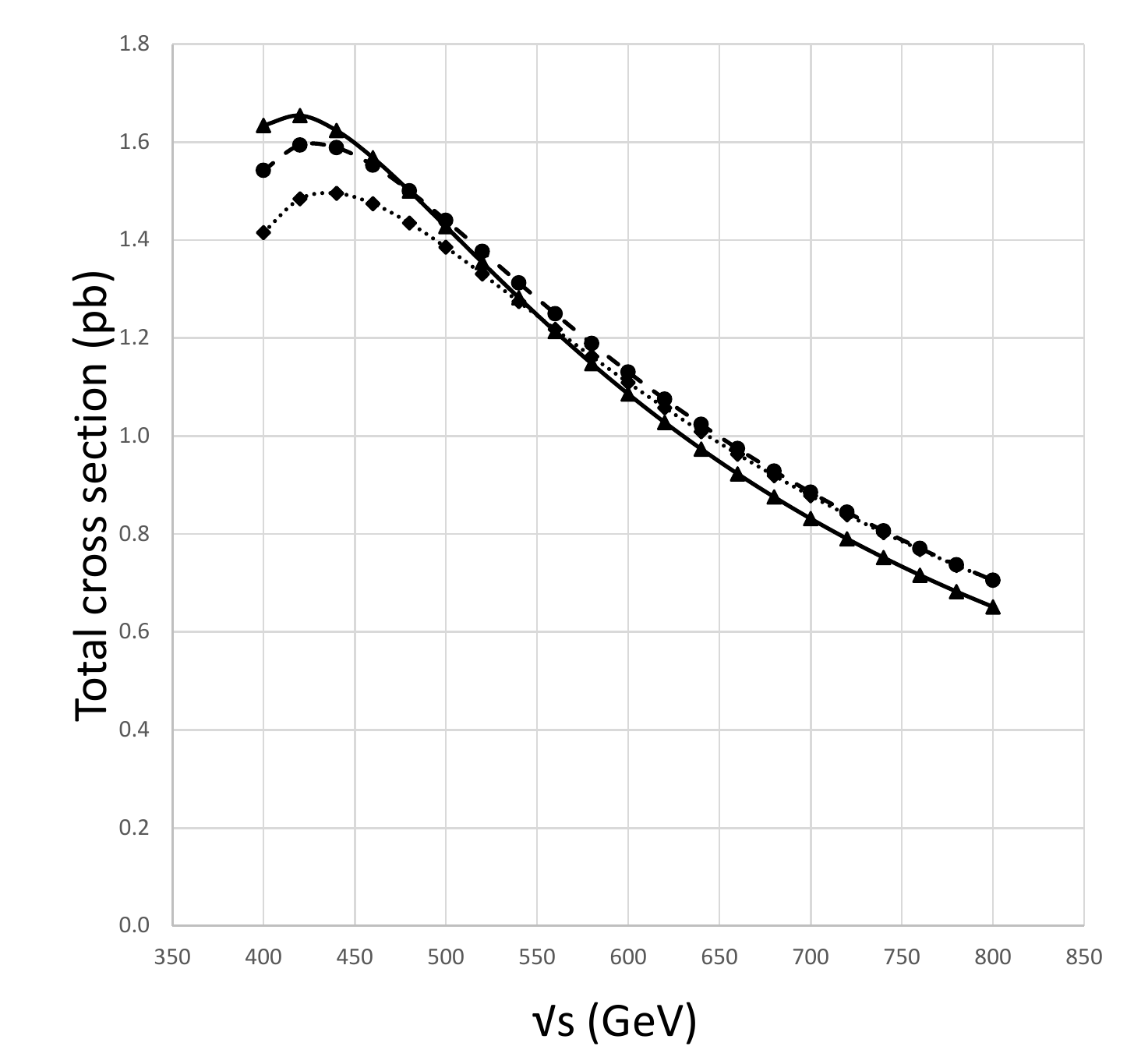}
  	\end{center}
  	\caption{
  	 Total cross sections of top quark pair production at the tree level, tree with the ISR and one-loop with the ISR corrections are shown as a function of the CM energy.
The solid line (with triangle points) and dashed line (with circle points) show the tree and NLO cross sections with the ISR.
The dotted line (with rectangle points) shows the cross sections with only the ISR correction on the tree cross sections.
}
  	\label{fig1}
  \end{figure}

At first, the fixed order correction without using an exponentiation method is investigated. 
Total cross sections obtained at leading (tree) and next-to-leading order (NLO) calculations are shown in Figure~\ref{fig1}.
The NLO calculations near the top-quark production threshold (around the CM energy of $400$ GeV) shows negative corrections about $7\%$.
This effect is called ``coulomb correction'', and it can be large because produced particles are moving slowly and have enough time-duration to interact with each other.
The corrections become very small around the CM energy of $500$ GeV and increase at the high energy region as to $2.8\%$ at the CM energy of $800$ GeV.
Among several types of radiative corrections, e.g., the initial and final state photon radiations, the vertex and box correction, and so on, the initial-state photonic correction give the largest contribution at the high energy region.
As shown in Figure~\ref{fig1}, the cross sections at tree level including the ISR correction (a dotted line in the figure) are almost the same as the full order $\mathcal{O}(\alpha)$ electroweak correction (a dashed line in the figure) at CM energies above $700$ GeV.
This means that the main contribution of the higher order corrections is caused by the initial state photonic corrections.
On the other hand, other corrections from the loop diagrams also give a large correction near the threshold region.
Around the CM energy $500$ GeV, these effects are cancelled accidentally and give a small correction on the total cross section.

\begin{figure}[t]
  	\begin{center}
  		\includegraphics[width={8cm}]{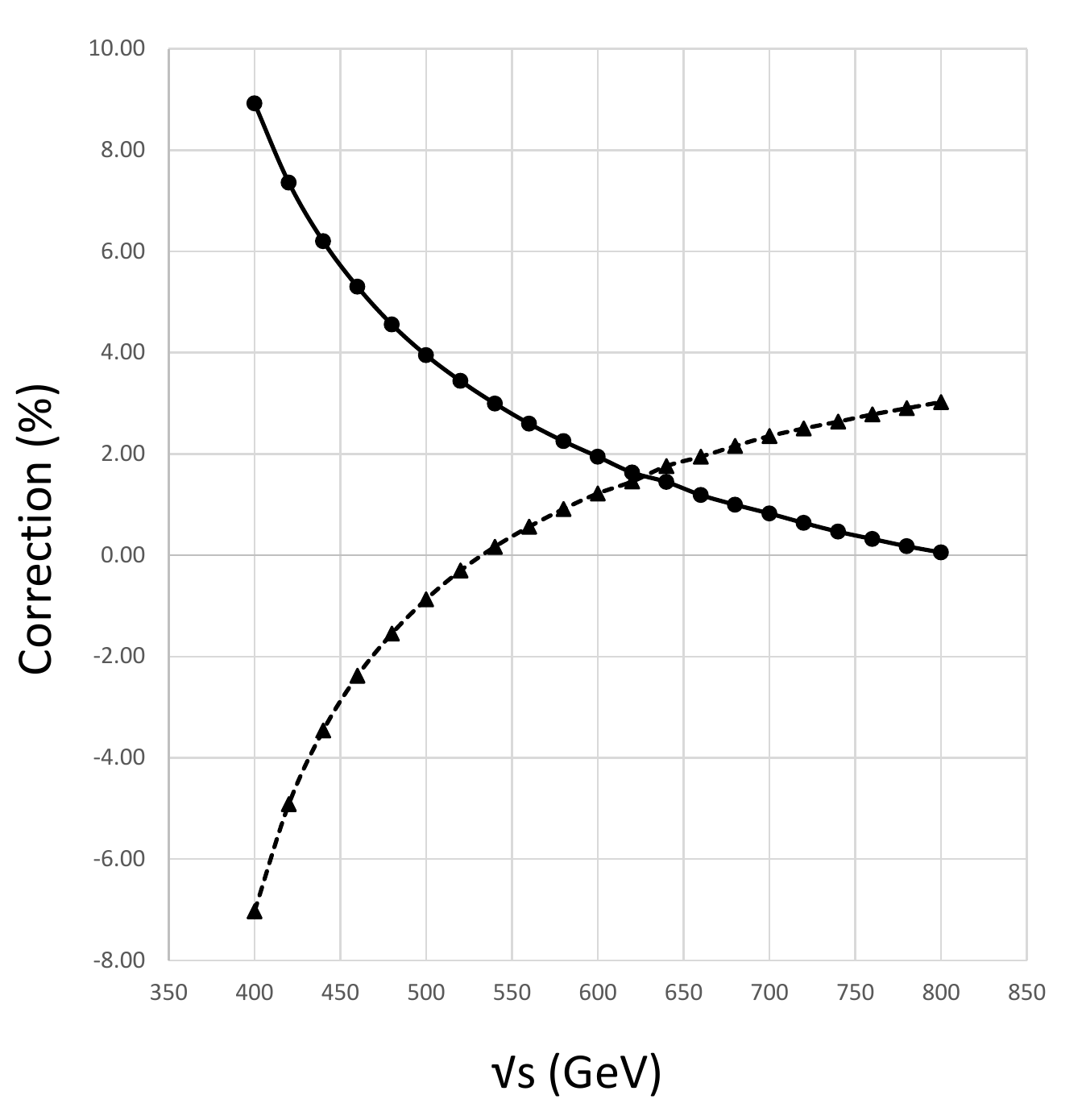}
  	\end{center}
  	\caption{
  	The correction degree of the full $\mathcal{O}(\alpha)$ electroweak radiative corrections (dashed line with triangle points) and the weak corrections (solid line with circle points).
  	}
  	\label{fig3}
  \end{figure}

After subtracting a trivial ISR correction from the total corrections, one can discuss the ``pure'' weak correction in the full $\mathcal{O}(\alpha)$ electroweak radiative corrections.
While the NLO correction degree is defined as
\begin{eqnarray}
\delta_{NLO}&=&\frac{\sigma_{NLO;fixed}-\sigma_{Tree}}{\sigma_{Tree}},
\end{eqnarray}
the weak correction degree is defined as
\begin{eqnarray}
\delta_{weak}&=&\frac{\sigma_{NLO;ISR}-\sigma_{ISR}}{\sigma_{ISR}}.
\end{eqnarray}
In the definition of $\delta_{weak}$, the trivial initial-state photonic corrections are subtracted from the full $\mathcal{O}(\alpha)$ electroweak radiative corrections, and thus, $\delta_{weak}$ shows a fraction of the mainly weak-correction in the full $\mathcal{O}(\alpha)$ electroweak radiative corrections.
The behaviors of $\delta_{weak}$ and $\delta_{NLO}$ are shown in Figure~\ref{fig3} with respect to the CM energies.
One can see that the weak correction becomes smaller and smaller at a high energy region, and arrives at almost zero at the CM energy of $800$ GeV.
On the other hand, at the CM energy of $500$ GeV, the pure-weak corrections gives $+4\%$ correction over the trivial ISR corrections.

\subsubsection{angular distribution}
  \begin{figure}[t]
  	\begin{center}
  		\includegraphics[width={8cm}]{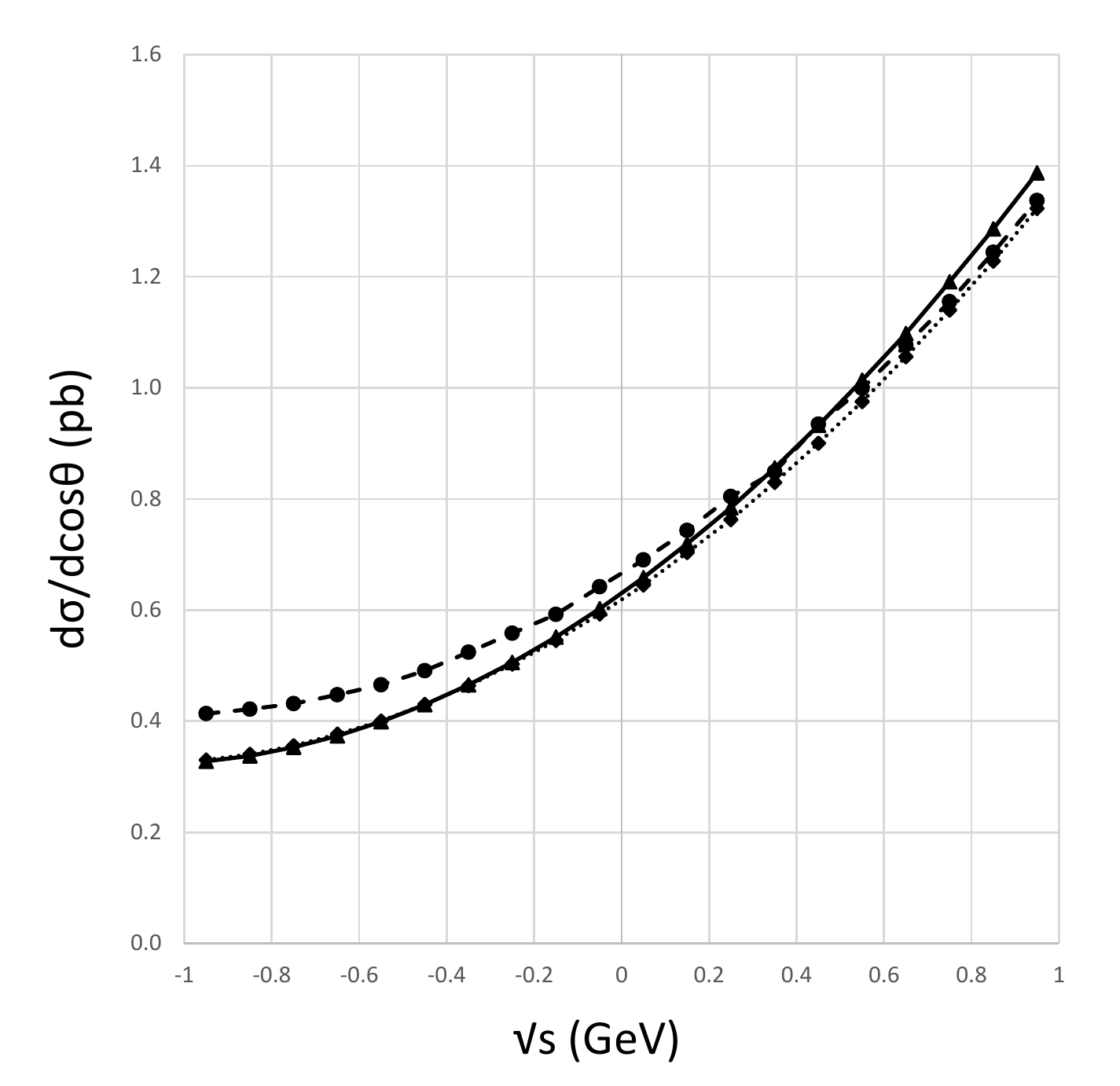}
  	\end{center}
  	\caption{
Angular distributions of a top quark.
A solid line (with triangle points), dotted line (with rectangle points)  and dashed line (with circle points) show the tree, tree with ISR and NLO with ISR cross sections.
  	}
  	\label{fig4}
  \end{figure}
As mentioned above, the radiative correction does not change the total cross section accidentally at the CM energy $500$ GeV.
While the total cross section stays the same after the radiative correction, an angular distribution is not the case.
In general, the real-photon emission affects the angular distribution such that a steep peak being mild. 
In reality, the ISR correction (a dotted line with rectangle points) makes a forward peak of a top quark production at a tree level (a solid line with triangle points) smaller as shown in Figure \ref{fig4}.
In addition to that, the weak correction raises the backward scattering as shown by a dashed line with circle points in Figure \ref{fig4}.
As the result, the total cross section does not change so much.

\subsection{Photonic correction at two-loop order}
\begin{figure}[t]
  	\begin{center}
  		\includegraphics[width={8cm}]{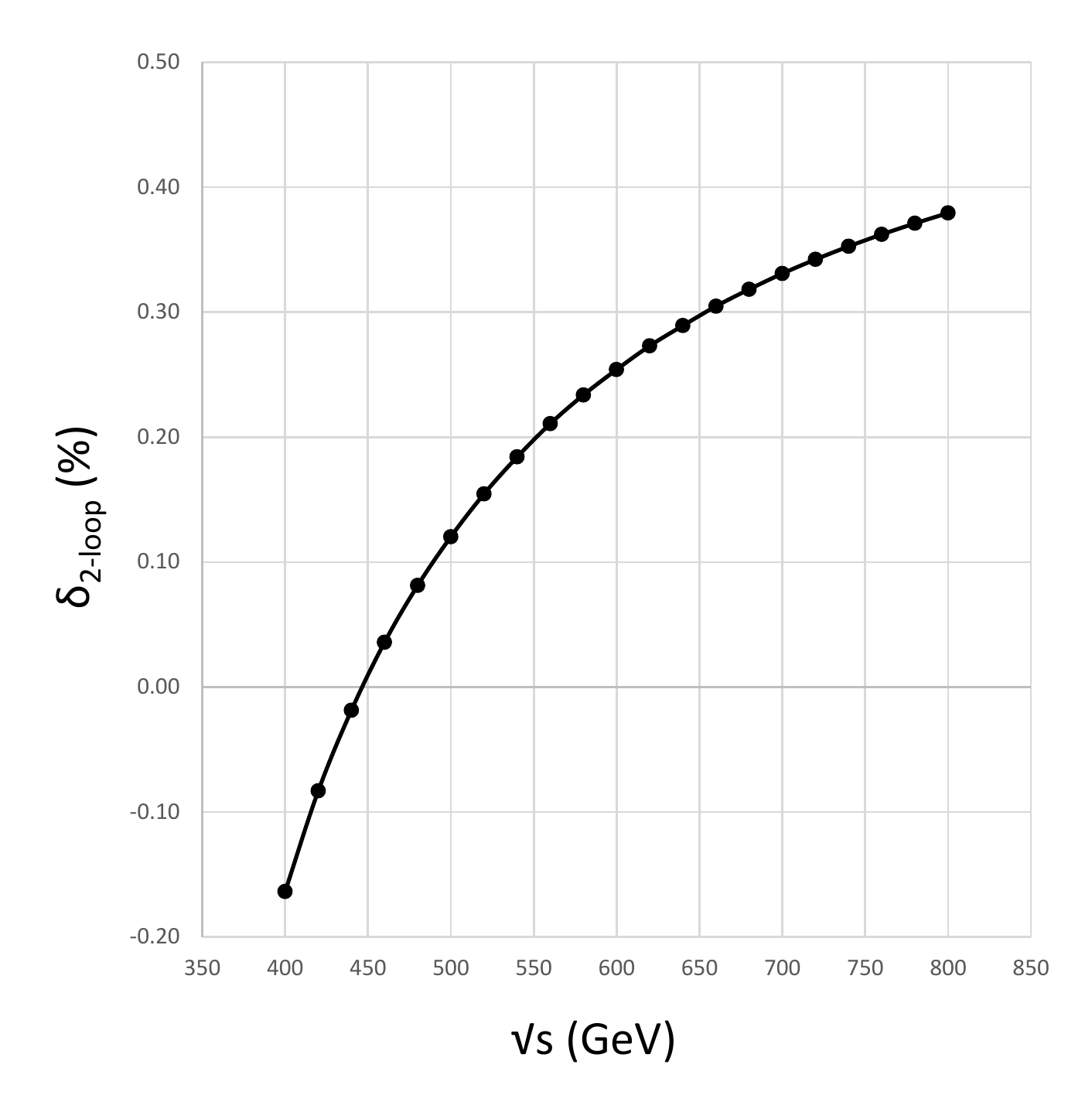}
  	\end{center}
  	\caption{
  	The contribution of the two-loop diagrams on the ISR corrections are show as a function of the CM energies.
  	The definition of  $\delta_{2\rm{-} loop}$ is given in (\ref{d2l}).
  	}
  	\label{fig5}
  \end{figure}

The structure function $H(s,x)$ given in (\ref{ISRLoop}) are including two-loop effects.
All of above results are obtained using full formula (\ref{ISRLoop}).
As mentioned in previous subsection,  a main contribution of the radiative corrections comes from the initial state photonic-correction at the high energy region.
Therefore, the ISR correction is important among many terms of the radiative corrections around these energies.
If the two-loop contribution has a significant fraction in the full correction, even higher-loop corrections must be considered for the future experiments.
The fraction of two-loop contribution over the one-loop one is defined as
\begin{eqnarray}
\delta_{2-loop}&=&\frac{\sigma_{ISR}-\sigma^{(1)}_{ISR}}{\sigma^{(1)}_{ISR}},\label{d2l}
\end{eqnarray}
where $\sigma^{(1)}_{ISR}$ shows the ISR corrected cross sections using the structure function (\ref{ISRLoop}) with omitting $\delta_2$ and $\beta^2$ terms.
Numerical results are shown in Figure~\ref{fig5}.
The two-loop contribution is smaller than $0.4\%$ in the energy region between $400$ GeV to $800$ GeV as shown in Figure~\ref{fig5}.

\subsection{Running coupling}
\begin{figure}[t]
  	\begin{center}
  		\includegraphics[width={8cm}]{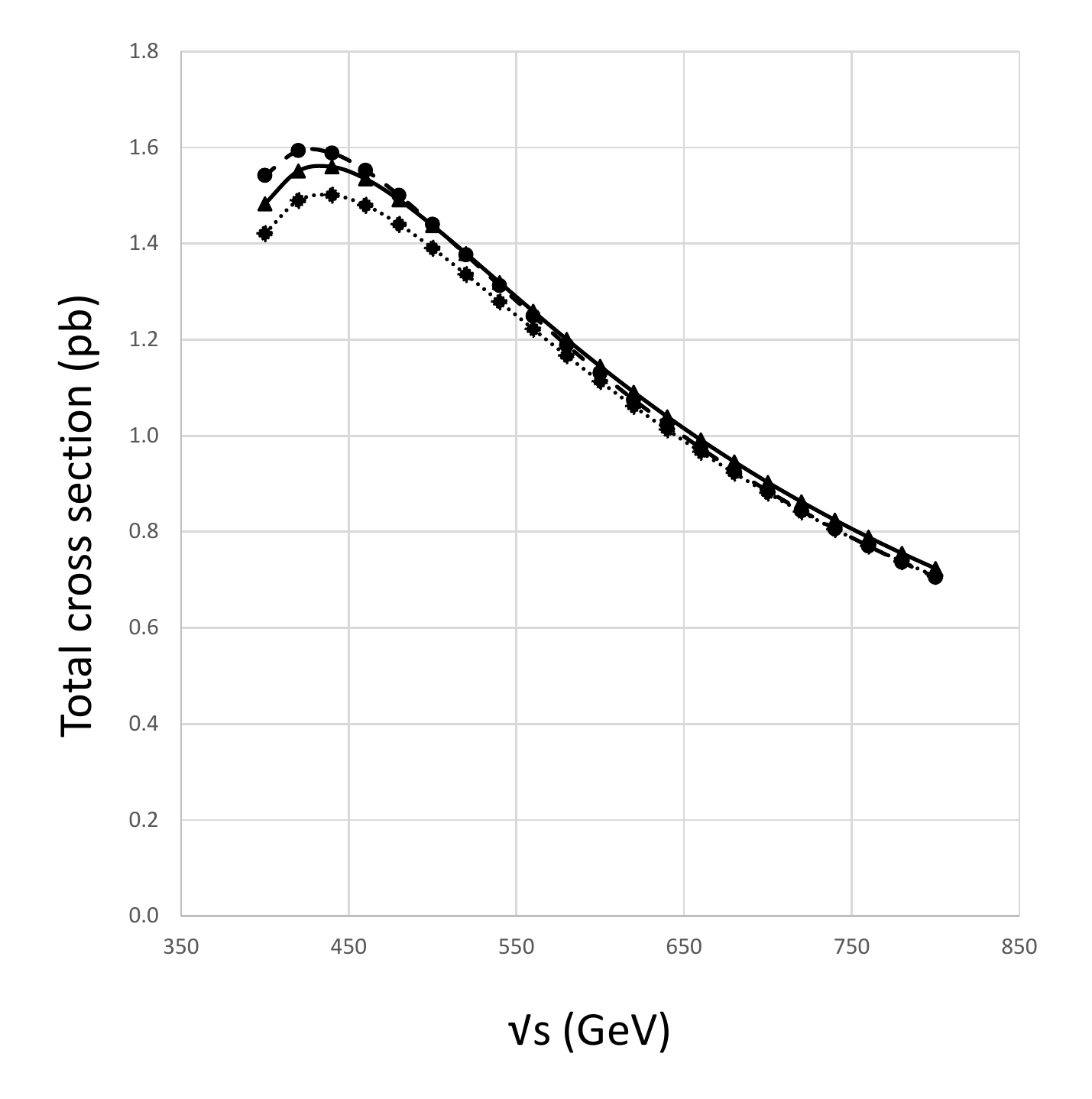}
  	\end{center}
  	\caption{
Total cross sections of top quark pair production of the tree and NLO cross sections with the ISR correction, and the improved Born cross sections are shown as a function of the CM energy.
The dotted line (with rectangle points) and dashed line (with circle points) show the tree and NLO cross sections with the ISR.
The solid line (with triangle points) shows the cross sections of the improved Born approximation.
  	}
  	\label{fig6}
  \end{figure}

Yet another improvement of the cross section estimation is known as a running coupling method.
This method is also taking a higher order effect of vacuum polarization diagrams into account as an effective coupling constant.
After summing up contributions from fermion-loops on gauge-boson propagators, those form solely a gauge-invariant subset, an electro-weak coupling effectively varies according to the four-momentum square of propagators.
The effective coupling $\alpha(s)$ at the energy scale $s$ can be written as;
\begin{eqnarray}
\alpha\left(|q^2|\right)&=&\frac{\alpha\left(\mu^2\right)}
{1-\frac{\alpha\left(\mu^2\right)}{3\pi}\log{\left(\frac{|q^2|}{\mu^2}\right)}},
\end{eqnarray}
where $q^2$ is the four-momentum square of a propagator of the target process.
At the same time, the weak mixing angle $\theta_W$ is obtained from the Fermi weak-coupling constant $G_F$ as\cite{Altarelli:473529};
\begin{eqnarray}
\sin^2{\theta_W}&=&\frac{\pi\alpha\left(|q^2|\right)}{\sqrt{2}G_Fm_W^2}.
\end{eqnarray}  
The improved cross section $\sigma_{imp}$ can be written as;
\begin{eqnarray}
\sigma_{imp}&=&\int^1_0 dx_1\int^{1-x_1}_0 dx_2~D(x_1,s)D(x_2,s)\sigma_{Tree}\left(\alpha\left(sx_1x_2\right);sx_1x_2\right),
\end{eqnarray}
where $\sigma_{Tree}\left(\alpha;s\right)$ is the tree cross section at the CM energy-square $s$ with the coupling constant $\alpha$.
This method is referred to as the improved Born approximation.
Numerical results are summarized in Figure \ref{fig6}.
Here the measured value of $\alpha(\mu^2=(2m_W)^2)=128.07$\cite{Altarelli:473529} is used. 
Above the CM energy of $500$ GeV, the improved Born method gives approximated values better than $2\%$ with respect to the NLO cross sections with the ISR.
%
%
\section{Summary}\label{Summary}
We calculated the precise cross sections of an $e^+e^-\rightarrow t\bar{t}$ process at an energy region from 400 GeV to 800 GeV.
Especially, the initial-state photon emissions are discussed in details. 
An exponentiation technique is applied for the initial-state photon emission up to two-loop order.
We found that the total cross section of  a top quark pair-production at a center of mass energy of 500 GeV receives  the weak corrections of $+4\%$ over the trivial ISR corrections.
Among the ISR contributions, two-loop diagrams gives less than $0.4\%$ correction with respect to the one-loop ones at the CM energies from $400$ GeV to $800$ GeV.

The improved Born approximation gives cross sections better than $2\%$ compared with the NLO cross sections with the ISR.

%
\bibliography{tT_ISR}

\providecommand{\href}[2]{#2}\begingroup\raggedright\begin{thebibliography}{10}

\bibitem{Aad:2012tfa}
{\scshape ATLAS} collaboration, G.~Aad et~al., \emph{{Observation of a new
  particle in the search for the Standard Model Higgs boson with the ATLAS
  detector at the LHC}},
  \href{https://doi.org/10.1016/j.physletb.2012.08.020}{\emph{Phys. Lett.}
  {\bfseries B716} (2012) 1--29}.

\bibitem{Chatrchyan201230}
{\scshape CMS} collaboration, S.~Chatrchyan et~al., \emph{{Observation of a new
  boson at a mass of 125 GeV with the CMS experiment at the LHC}},
  \href{https://doi.org/10.1016/j.physletb.2012.08.021}{\emph{Phys. Lett.}
  {\bfseries B716} (2012) 30--61}.

\bibitem{Behnke:2013xla}
T.~Behnke, J.~E. Brau, B.~Foster, J.~Fuster, M.~Harrison, J.~M. Paterson
  et~al., \emph{{The International Linear Collider Technical Design Report -
  Volume 1: Executive Summary}},
  \href{https://arxiv.org/abs/1306.6327}{{\ttfamily 1306.6327}}.

\bibitem{Baer:2013cma}
H.~Baer, T.~Barklow, K.~Fujii, Y.~Gao, A.~Hoang, S.~Kanemura et~al., \emph{{The
  International Linear Collider Technical Design Report - Volume 2: Physics}},
  \href{https://arxiv.org/abs/1306.6352}{{\ttfamily 1306.6352}}.

\bibitem{Fujimoto:1987hu}
J.~Fujimoto and Y.~Shimizu, \emph{{Radiative Corrections to $e^+ e^-
  \rightarrow t\bar{t}$ in Electroweak Theory}},
  \href{https://doi.org/10.1142/S0217732388000696}{\emph{Mod. Phys. Lett.}
  {\bfseries 3A} (1988) 581}.

\bibitem{Fleischer:2002rn}
J.~Fleischer, T.~Hahn, W.~Hollik, T.~Riemann, C.~Schappacher and
  A.~Werthenbach, \emph{{Complete electroweak one loop radiative corrections to
  top pair production at TESLA: A Comparison}},
  \href{https://arxiv.org/abs/hep-ph/0202109}{{\ttfamily hep-ph/0202109}}.

\bibitem{Fleischer2003}
J.~Fleischer, A.~Leike, T.~Riemann and A.~Werthenbach, \emph{Electroweak
  one-loop corrections for $e^+e^-$- annihilation into $t\bar{t}$ including
  hard bremsstrahlung},
  \href{https://doi.org/10.1140/epjc/s2003-01263-8}{\emph{The European Physical
  Journal C - Particles and Fields} {\bfseries 31} (2003) 37--56}.

\bibitem{Khiem2013}
P.~H. Khiem, J.~Fujimoto, T.~Ishikawa, T.~Kaneko, K.~Kato, Y.~Kurihara et~al.,
  \emph{Full o$(\alpha)$ electroweak radiative corrections to
  $e^+e^-\rightarrow t\bar{t}\gamma$ with grace-loop},
  \href{https://doi.org/10.1140/epjc/s10052-013-2400-3}{\emph{The European
  Physical Journal C} {\bfseries 73} (2013) 2400}.

\bibitem{Quach:2017ijt}
N.~M.~U. Quach, Y.~Kurihara, K.~H. Phan and T.~Ueda, \emph{{Beam polarization
  effects on top-pair production at the ILC}},
  \href{https://arxiv.org/abs/1706.03432}{{\ttfamily 1706.03432}}.

\bibitem{doi:10.1093/ptep/ptx048}
Y.~Kouda, T.~Kon, Y.~Kurihara, T.~Ishikawa, M.~Jimbo, K.~Kato et~al., \emph{One
  loop effects of natural susy in third generation fermion production at the
  ilc}, \href{https://doi.org/10.1093/ptep/ptx048}{\emph{Progress of
  Theoretical and Experimental Physics} {\bfseries 2017} (2017) 053B02}.

\bibitem{Belanger2006117}
G.~B\'{e}langer, F.~Boudjema, J.~Fujimoto, T.~Ishikawa, T.~Kaneko, K.~Kato
  et~al., \emph{Automatic calculations in high energy physics and grace at
  one-loop},
  \href{https://doi.org/http://dx.doi.org/10.1016/j.physrep.2006.02.001}{\emph{Physics
  Reports} {\bfseries 430} (2006) 117 -- 209}.

\bibitem{PhysRevD.75.113002}
J.~Fujimoto, T.~Ishikawa, Y.~Kurihara, M.~Jimbo, T.~Kon and M.~Kuroda,
  \emph{Two-body and three-body decays of charginos in one-loop order in the
  mssm}, \href{https://doi.org/10.1103/PhysRevD.75.113002}{\emph{Phys. Rev. D}
  {\bfseries 75} (Jun, 2007) 113002}.

\bibitem{Belanger2003152}
G.~B\'{e}langer, F.~Boudjema, J.~Fujimoto, T.~Ishikawa, T.~Kaneko, Y.~Kurihara
  et~al., \emph{Full $\cal{O}$$(\alpha)$ electroweak corrections to double
  higgs-strahlung at the linear collider},
  \href{https://doi.org/http://dx.doi.org/10.1016/j.physletb.2003.09.080}{\emph{Physics
  Letters B} {\bfseries 576} (2003) 152 -- 164}.

\bibitem{Belanger2003163}
G.~B\'{e}langer, F.~Boudjema, J.~Fujimoto, T.~Ishikawa, T.~Kaneko, K.~Kato
  et~al., \emph{Full $\cal{O}$$(\alpha)$ electroweak and $\cal{O}$$(\alpha_s)$
  corrections to $e^+e^-\rightarrow t\bar{t}h$},
  \href{https://doi.org/http://dx.doi.org/10.1016/j.physletb.2003.07.072}{\emph{Physics
  Letters B} {\bfseries 571} (2003) 163 -- 172}.

\bibitem{Belanger2003353}
G.~B\'{e}langer, F.~Boudjema, J.~Fujimoto, T.~Ishikawa, T.~Kaneko, K.~Kato
  et~al., \emph{Full $\cal{O}$$(\alpha)$ corrections to $e^+e^-\rightarrow
  \nu\bar{\nu}h$ by grace},
  \href{https://doi.org/http://dx.doi.org/10.1016/S0920-5632(03)80198-6}{\emph{Nuclear
  Physics B - Proceedings Supplements} {\bfseries 116} (2003) 353 -- 357}.

\bibitem{Kato:2005iw}
K.~Kato, F.~Boudjema, J.~Fujimoto, T.~Ishikawa, T.~Kaneko, Y.~Kurihara et~al.,
  \emph{{Radiative corrections for Higgs study at the ILC}}, {\emph{PoS}
  {\bfseries HEP2005} (2006) 312}.

\bibitem{doi:10.1143/PTPS.73.1}
K.-i. Aoki, Z.~Hioki, R.~Kawabe, M.~Konuma and T.~Muta, \emph{Electroweak
  theoryframework of on-shell renormalization and study of higher-order
  effects}, \href{https://doi.org/10.1143/PTPS.73.1}{\emph{Progress of
  Theoretical Physics Supplement} {\bfseries 73} (1982) 1}.

\bibitem{doi:10.1143/PTPS.100.1}
J.~Fujimoto, M.~Igarashi, N.~Nobuya, S.~Yoshimitsu and T.~Keijiro,
  \emph{Radiative corrections to $e^+e^-$ reactions in electroweak theory},
  \href{https://doi.org/10.1143/PTPS.100.1}{\emph{Progress of Theoretical
  Physics Supplement} {\bfseries 100} (1990) 1}.

\bibitem{Vermaseren:2000nd}
J.~A.~M. Vermaseren, \emph{{New features of FORM}},
  \href{https://arxiv.org/abs/math-ph/0010025}{{\ttfamily math-ph/0010025}}.

\bibitem{vanOldenborgh:1990yc}
G.~J. van Oldenborgh, \emph{{FF: A Package to evaluate one loop Feynman
  diagrams}}, \href{https://doi.org/10.1016/0010-4655(91)90002-3}{\emph{Comput.
  Phys. Commun.} {\bfseries 66} (1991) 1--15}.

\bibitem{Hahn:1998yk}
T.~Hahn and M.~Perez-Victoria, \emph{{Automatized one loop calculations in
  four-dimensions and D-dimensions}},
  \href{https://doi.org/10.1016/S0010-4655(98)00173-8}{\emph{Comput. Phys.
  Commun.} {\bfseries 118} (1999) 153--165}.

\bibitem{Kawabata1986127}
S.~Kawabata, \emph{A new monte carlo event generator for high energy physics},
  \href{https://doi.org/http://dx.doi.org/10.1016/0010-4655(86)90025-1}{\emph{Computer
  Physics Communications} {\bfseries 41} (1986) 127 -- 153}.

\bibitem{KAWABATA1995309}
S.~Kawabata, \emph{A new version of the multi-dimensional integration and event
  generation package bases/spring},
  \href{https://doi.org/http://dx.doi.org/10.1016/0010-4655(95)00028-E}{\emph{Computer
  Physics Communications} {\bfseries 88} (1995) 309 -- 326}.

\bibitem{Boudjema:1995cb}
F.~Boudjema and E.~Chopin, \emph{{Double Higgs production at the linear
  colliders and the probing of the Higgs selfcoupling}},
  \href{https://doi.org/10.1007/s002880050298}{\emph{Z. Phys.} {\bfseries C73}
  (1996) 85--110}.

\bibitem{Khiem2015192}
P.~Khiem, Y.~Kurihara, J.~Fujimoto, T.~Ishikawa, T.~Kaneko, K.~Kato et~al.,
  \emph{Full electroweak radiative corrections to at the ilc with grace-loop},
  \href{https://doi.org/http://dx.doi.org/10.1016/j.physletb.2014.11.048}{\emph{Physics
  Letters B} {\bfseries 740} (2015) 192 -- 198}.

\bibitem{Olive:2016xmw}
{\scshape Particle Data Group} collaboration, C.~Patrignani et~al.,
  \emph{{Review of Particle Physics}},
  \href{https://doi.org/10.1088/1674-1137/40/10/100001}{\emph{Chin. Phys.}
  {\bfseries C40} (2016) 100001}.

\bibitem{Altarelli:473529}
CERN, \emph{{Workshop on Physics at LEP2, v.2}}, (Geneva), CERN, 1996.

\end{thebibliography}\endgroup
\end{document}